\documentstyle[aps,epsf,preprint]{revtex}
\tighten

\begin{document}
\draft

\title{Production of bound {$\mu^{+}\mu^{-}$}-systems
in relativistic heavy ion collisions}

\author{I.~F.~Ginzburg$^{1,2,}$\thanks{E-mail: ginzburg@math.nsc.ru},
U.~D.~Jentschura$^{3,}$\thanks{E-mail: ulrich@theory.phy.tu-dresden.de},
S.~G.~Karshenboim$^{2,4,}$\thanks{E-mail: sgk@onti.vniim.sbp.su},
F.~Krauss$^{3,}$\thanks{E-mail: krauss@theory.phy.tu-dresden.de},\\
V.~G.~Serbo$^{2,5,}$\thanks{E-mail: serbo@math.nsc.ru} and
G.~Soff$^{3,}$\thanks{E-mail: soff@physik.tu-dresden.de}}

\author{~}

\address{$^{1}$Institute for Mathematics, 630090 Novosibirsk, Russia}
\address{$^{2}$Max--Planck--Institut f\"{u}r Physik komplexer Systeme,\\
Bayreuther Stra\ss e 40, 01069 Dresden, Germany\thanks{temporary address}}
\address{$^{3}$Institut f\"ur Theoretische Physik, TU Dresden, 01062
Dresden, Germany}
\address{$^{4}$D. I. Mendeleev Institute for Metrology,
198005 St.~Petersburg, Russia}
\address{$^{5}$Novosibirsk State University, 630090 Novosibirsk, Russia}

\maketitle

\begin{abstract}
Dimuonium (the bound system of two muons, $\mu^+\mu^-$-atom) has not been
observed yet. In this paper we discuss the electromagnetic production of 
dimuonium at RHIC and LHC in relativistic heavy ion collisions. 
The production of parastates is analyzed in the 
equivalent photon approximation. For the treatment of orthostates,
we develop a three photon formalism. We determine the production rates  
at RHIC and LHC with an accuracy of a few percent
and discuss problems related to the observation of dimuonium.
\end{abstract}


\section{INTRODUCTION}
\label{intro}

The study of exotic electromagnetic bound systems and their 
properties is of theoretical and experimental interest.
The bound $\mu^+\mu^-$ system (dimuonium, DM) has been subject to 
extensive theoretical investigations 
\cite{CL,nlo,nemenov,malenfant,Bilenkii}.
As demonstrated in \cite{nlo},
the decay rate of the dimuonic system is sensitive to 
radiative corrections from the so far unexplored time--like region of QED. 

Although dimuonium has not been observed yet, 
a lot of different pathways for its 
production have been considered. For example, the production of dimuonium
in the decay of the $\eta$-meson ($\eta\to {\rm DM} + \gamma$) 
was investigated in Refs.~\cite{nemenov,Kozlov},
and in Ref.~\cite{malenfant} the decay $K_L^0 \to {\rm DM} + \gamma$ 
was considered. It has to be mentioned that in decays it is
possible to produce only the $S=1$ orthostates of dimuonium. 
Other calculations were performed for the 
production of dimuonium in collisions of charged particles
(see Refs. \cite{Bilenkii,Holvik}) and 
in collisions of photons with nuclei \cite{Bilenkii}. 

In this paper we investigate quite a different mechanism, which is based on 
the availability of relativistic heavy ions at high luminosities. Two new 
large hadron colliders, RHIC and LHC, are scheduled to be operative for the 
next decade. In Table \ref{colli} we list the decisive experimental 
parameters of the new colliders (see Refs.~\cite{revpart,peggs,brandt}).
We consider here the purely electromagnetic production channel
\[
A_1\,A_2 \to A_1\,A_2 + {\rm DM}\,,
\]
where the $A_i$ represent relativistic nuclei with nuclear charge
numbers $Z_i$, and DM stands both for the $S=0$ parastates of dimuonium
(paradimuonium, PM) and for the $S=1$ orthostates (orthodimuonium, OM). 
Because the nuclei do not change during the production 
process, they emit the photons
{\em coherently}. This means that the perturbation 
parameter associated with each photon exchange between the nuclei and
and the produced system is not $\alpha\approx1/137$,
but rather $Z\alpha\sim 0.6$ (for Au and Pb). This
leads to a very large flux of equivalent photons available
for the production of exotic particles.

The $C$-even PM can be produced in collisions of an even number of 
virtual photons (two photon production mechanism,
see the diagram of Fig.~\ref{prod}{\sl a}. 
The $C$-odd orthostate (OM) can only 
be produced by an odd number of virtual photons, i.e. via 
bremsstrahlung production (one photon, Fig.~\ref{prod}{\sl b}) and 
three photon production (see Fig.~\ref{prod}{\sl c}). 
We consider here mainly the production of PM by two photons and the 
production of OM by three photons. Two photon
and three photon fusion is the dominating process for the production
of parastates and orthostates, respectively.
The influence of multiphoton processes on the production
rate is described by the effective perturbation parameter 
\begin{equation}\label{rho1}
\rho = \left(\frac{Z\alpha\Lambda}{m_{\mu\mu}}\right)^2
\stackrel{<}{\sim} 0.04\,,
\end{equation}
with 
\begin{equation}\label{Lambda}
1/\Lambda^2 = 1/6 \, \langle r^2 \rangle\,,
\end{equation}
where $\langle r^2 \rangle$ is the mean square radius 
of the charge distribution of the nucleus, and the mass of the
dimuonic atom is $m_{\mu\mu}\approx 2\,m_\mu = 211\,\mbox{ MeV}$.
Therefore, in all cases under consideration the multiphoton processes set 
limits on the accuracy on the level of $5\,\%$. 

For our purpose it is sufficient to treat the dimuonia as compound neutral 
particles. To a good approximation their production rate then is proportional 
to the square of the wave function at the origin. In non--relativistic 
approximation it is only the probability density of S states at the origin 
which does not vanish,
\begin{equation}
\left \vert \psi_{\rm nS}(0) \right \vert^2 =
\frac{\alpha^3\,m_{\mu}^3}{8\pi\,n^3} \,.
\label{width}
\end{equation}
The production rate and the lifetime of the dimuonic 
atoms are both proportional to this value.
The lifetimes of low lying states are of the 
order of $\tau\sim 10^{-12}\,{\rm s}$
and are summarized in Table~\ref{restable}. 
A brief discussion of the evaluation of the 
lifetime of parastates is given in Appendix \ref{app:a}. 
Main decay channels are the annihilation processes
\begin{equation}
\label{decay}
{\rm PM}\to \gamma\gamma \,,\quad {\rm OM}\to e^+e^-\,.
\end{equation}
The rate of atomic transitions from excited S states to lower
atomic states is of the same order of magnitude ($\alpha^5\,m$) as the
annihilation decay rate. It results in additional final states
via atomic decays of excited DM levels which cascade through
${\rm S} \to {\rm P} \to {\rm S}$ transitions.
This leads to observable X-ray photons (at least two quanta)
having ``atomic'' energy $\sim \alpha^2\,m_\mu\,(n'{}^{-2}-n^{-2})/4$.
Main properties of the various states can be found in Table~\ref{restable}
together with 2P paradimuonium which is produced in atomic transitions
from 3S and 4S.

The detection of dimuonium would constitute a continuation of 
the recent investigations of exotic bound systems. Over the past years,
experiments on antihydrogen \cite{Munger,Baur}, pionium \cite{Afanasyev93}
and the bound $\pi\mu$ system \cite{Coombes,Aronson} have been reported.

This paper is organized as follows: first we investigate the production
of paradimuonium in Sec.~\ref{secpara}. We then proceed to orthodimuonium,
which is discussed in Sec.~\ref{secortho}. Finally we discuss the background
in Sec.~\ref{secback} and summarize the results in Sec.~\ref{secres}.

\section{PARADIMUONIUM PRODUCTION}
\label{secpara}

\begin{figure}[htb]
\centerline{\mbox{\epsfysize=5.0cm\epsffile{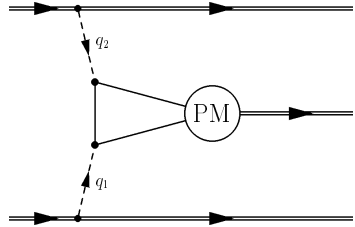}}}
\caption{\label{twophoton} Two photon production of
paradimuonium by relativistic heavy nuclei.}
\end{figure}

The production of an $S=0$ parastate of dimuonium by a two photon process
is represented by the diagram in Fig.~\ref{twophoton}.
The diagram is evaluated using the equivalent photon approximation in the 
approach originally presented in Ref.~\cite{BGMS}. Two nuclei $A_1$ and $A_2$ 
with identical charge number $Z$ and atomic mass number $A$ colliding with 
each other emit $dn_i\,,(i=1,2)$ equivalent virtual photons within the 
energy ranges $(\omega_i, \omega_i + d\omega_i)$ and with four-momenta 
denoted as $q_i$. The virtualities of the photons are $Q_i^2=-q_i^2$. Upon 
fusion, these photons produce a PM-bound state with four-momentum 
$p=q_1+q_2$. Its mass squared $p^2 = W^2 = (q_1 +q_2)^2$ is approximately 
equal to $4\,m_{\mu}^2$. The most important contribution to the production 
process stems from photons with very small virtualities 
$Q_i^2 \ll m_{\mu}^2$. To a good approximation, the photons move in 
opposite directions, and we have $W^2 \approx 4\,\omega_1\,\omega_2$. In this 
very region the differential cross section $d\sigma$ for the 
$A_1\,A_2 \to A_1\,A_2 + {\rm PM}$ process is related to the cross section 
$\sigma_{\gamma \gamma}$ for the process $\gamma\gamma \to {\rm PM}$ by the 
equation
\begin{equation}
\label{dsigmaPM}
d\sigma_{{\rm{\scriptscriptstyle PM}}} = dn_1 dn_2 \, \sigma_{\gamma \gamma} (W^2)\,.
\end{equation}
The spectrum of equivalent photons is given by Eq. (D.4) in Ref.~\cite{BGMS}, 
which upon omission of terms of order $\omega_i /E \ll 1$ reads
\begin{equation}
\label{spectrum1}
dn_i(\omega_i,Q_i^2) = {Z^2 \alpha \over \pi}\, {d\omega_i \over \omega_i}\,
\left(1 - {Q^2_{i\,\min} \over Q_i^2 }\right)\,F^2(Q^2_i)
\, {dQ^2_i\over Q^2_i},\quad
Q^2_{i\,\min} = {\omega_i^2\over \gamma^2}.
\end{equation}
In the calculations below we do not use the exact form factor of the nucleus 
$F(Q^2)$ but a simple approximation. This approximation corresponds to an 
exponentially decreasing charge distribution of the nucleus, whose mean 
square radius is adjusted to fit the experimental value 
(see Ref. \cite{tsai}, Eq. (B49)).
\begin{equation}
\label{formapprox}
F(Q^2) = \frac{1}{1 + Q^2/\Lambda^2} \quad \mbox{where} \quad 
\Lambda^2 = {0.164\; {\rm GeV}^2 \over A^{2/3}}.
\end{equation}
According to Eq.~(\ref{Lambda}), for Pb and Au the parameter $\Lambda \approx 
70\,{\rm MeV}$, and for Ca $\Lambda\approx 118\,{\rm MeV}$. The approximate 
form factor enables us to perform some calculations analytically which else 
could only be done numerically. 

It is useful to note that the integral over $Q^2$ converges fast
at $Q^2 > \Lambda^2$. The decisive region of
integration is given by the condition 
$Q^2_{min}\le Q^2 \stackrel{<}{\sim}
\Lambda^2$ (cf.~Eq.~(\ref{spectrum1})). Therefore the main contribution
to the cross section is given by virtual photons with energies
\begin{equation}
\label{energy}
\omega_i \stackrel{<}{\sim} \Lambda \gamma\,.
\end{equation}

Because the two photon width of paradimuonium is small in comparison with its 
mass, we can use a $\delta$--approximation for the cross section 
$\sigma_{\gamma \gamma}$ (for further details see Eq.~(3.24) in 
Ref.~\cite{BGMS} and Eq.~(89.4) in Ref.~\cite{berestetskii}). For the 
$1 ^{1}S_0$ para ground state, this approximation has the form
\begin{equation}
\label{deltaapprox}
\sigma_{\gamma \gamma} (p^2) = 2\,\pi^2\,\alpha^5\,
\delta(p^2-4\,m_{\mu}^2)\,.
\end{equation}
After the transformation
\begin{eqnarray}
{d\omega_1 \over \omega_1}\; {d\omega_2 \over \omega_2}\;
\delta(p^2 -4\,m_{\mu}^2)&=&
{d\omega_1 \over \omega_1} \, {dp^2
\over p^2} \, \delta(p^2 -4\,m_{\mu}^2)
\nonumber\\
&\to& {1\over 4\,m_{\mu}^2}
\, {d\omega_1 \over \omega_1}
\end{eqnarray}
we cast the cross section into the form
\[
d\sigma_{\rm PM}= {\pi^2\over 2}\, \alpha^5\,
{d\omega_1\over \omega_1}\,{dn_1(\omega_1,Q_1^2)\over d\omega_1}
{dn_2(\omega_2,Q_2^2)\over d\omega_2}\,,
\]
where $\omega_2= m_\mu^2/ \omega_1$.  Using this formula, we derive the 
distribution of the produced PM atoms with respect to the energy 
$\varepsilon$ and the transverse momentum $p_\bot$ via the relations
\begin{equation}
\label{defenergy}
\varepsilon =\omega_1+{m_{\mu}^2\over\omega_1}\;\;\;,
\quad {\bf p}_\bot ={\bf q}_{1\,\bot} +{\bf
q}_{2\,\bot}\,.
\label{epsilon}
\end{equation}
It is useful to note that the integral over $Q^2$ converges fast for 
$Q^2 > \Lambda^2$. Integrating $dn_i(\omega ,Q_i^2)$ over $Q_i^2$, we obtain 
the equivalent photon spectrum in dependence on the energy, $dn_i(\omega)$.
\begin{equation}
\label{spectrum}
dn_i(\omega_i) = {Z^2 \alpha \over \pi} \,
f\left({\omega_i \over \Lambda \gamma}\right)\,
{d\omega_i\over \omega_i}\,.
\end{equation}
The function
\begin{equation}
\label{deffy}
f(x)= (1 + 2\,x^2)\, \ln{\left({1\over x^2} +1\right)}\, -\, 2
\end{equation}
drops very quickly at large $x$ in accordance with Eq.~(\ref{energy}) (indeed,
$f(x) \leq 1/(6\,x^4)$ at $x\geq 1$). Finally, we obtain
\begin{equation}
\label{sigmaPM}
\sigma_{\rm {\scriptscriptstyle PM}} = 
{Z^4\,\alpha^7 \over 2 \, m_{\mu}^2} \,G(\delta )\,,\quad \mbox{where} \quad
\delta = {m_{\mu} \over \Lambda \gamma }
\end{equation}
and
\begin{eqnarray}   \label{Gdelta}
G(\delta )&=&
\int\limits^{\omega_{\max}}_{\omega_{\min}} {d\omega_1 \over
\omega_1}\; f\left({\omega_1 \over \Lambda \gamma}\right)\,
f\left({\omega_2 \over \Lambda \gamma}\right)
\nonumber\\
&=&\int\limits^{x_{\max}}_{x_{\min}} {dx\over x} \,
f(x \, \delta) \, f(\delta/x)
\;, \end{eqnarray}
with $x =\omega_1/ m_\mu$. Because $\omega_i < E$  and 
$\omega_1 \omega_2 = m_\mu^2$ we have $x_{min} =m_\mu/E$ and 
$x_{max} =E/m_\mu$. However, due to the fast decline of $f(x)$ at $x>1$ we 
can expand these limits up to $x_{\min}= 0$ and $x_{\max} = \infty$ 
in very good approximation.

\begin{figure}[htb]
\centerline{\mbox{\epsfysize=7.0cm\epsffile{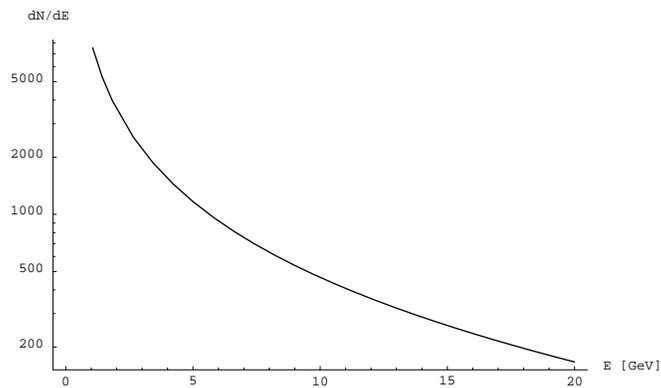}}}
\caption{\label{distrib} Distribution of paradimuonium produced 
at LHC in the Pb mode over the energy (in GeV). The distribution is normalized to the
annual production rate of $40\,000$ particles. The median of the
distribution is at $1.12\,{\rm GeV}$.}
\end{figure}

Numerical evaluation of the integral in Eq.~(\ref{Gdelta}) yields the 
following result for the total production cross sections
\begin{equation}
\label{resPM}
\sigma_{\rm{\scriptscriptstyle PM}} = 10^{-30}\; {\rm cm}^2 \times
\left\{\begin{array}{c l}
0.15  & \mbox{ for RHIC, Au mode}\,,\\
1.35  & \mbox{ for LHC, Pb mode}\,,\\
0.0066  & \mbox{ for LHC, Ca mode}\,.
\end{array}\right.
\end{equation}
The production cross sections for excited nS states are derived from the 
above cross section, which is obtained for the 1S states, with the aid of 
Eq.~(\ref{width}),
\begin{equation}
\label{sigmasn}
\sigma(nS) ={\sigma(1S)\over n^3}\,.
\end{equation}
The summation over $n$ enhances the result of Eq.~(\ref{resPM})
by a factor of
\begin{equation}
\zeta(3)=\sum_{n=1}^\infty {{1\over n^{3}}}\approx 1.202\,.
\label{zeta}
\end{equation}
The distribution in energy for the paradimuonium atoms is given by the 
integrand of Eq.~(\ref{Gdelta}), using the relation Eq.~(\ref{epsilon}).
It is shown in Fig.~\ref{distrib}.

We checked that the results depend only weakly on the choice of the form 
factor. With the help of a numerical computer program 
\cite{KGS} we took into account the 
Gaussian form factor $\exp(-Q^2/\Lambda_g^2)$ with 
$\Lambda_g = 60\,{\rm MeV}$ (for Pb and Au collisions) and
$\Lambda_g = 100\,{\rm MeV}$ (for Ca collisions) fixed on 
$\langle r^2\rangle$. We found that this changes the final result presented 
in Eq.~(\ref{resPM}) by less than one percent. 
The effect of omitting terms of the order ${\cal O}(Q_i^2/m_\mu^2)$
in the equivalent photon spectrum Eq.~(\ref{spectrum1}) is also
negligible. The relative contribution of the omitted terms is of the order of
\begin{equation}
\eta_2 = {\Lambda^2 \over 2 m_\mu^2 \, L}
\, ,\mbox{ and with } L=\ln{1\over \delta^2} =
\ln{{\Lambda^2 \gamma^2\over m_\mu^2}} \quad \mbox{it follows} \quad
\eta_2\sim (1\div 2)\% \,.
\end{equation}
The effect of processes involving more than two photons is governed by the 
parameter $\rho$ from Eq. (\ref{rho1}) and we recall that their effect is
on the level of roughly 5 \%. 

It is instructive to consider additionally the leading logarithmic 
approximation (LLA) for the process. In the LLA, we approximate $f(x)$ by 
$2\,\ln (1/ x)$.  The restriction $Q^2_{i\,\min} \stackrel{<}{\sim} \Lambda^2$
corresponds to $m^2_{\mu}/(\Lambda \gamma) < \omega_1 < \Lambda\gamma$. 
Therefore
\begin{equation}
G^{{\rm{\scriptscriptstyle LLA}}}(\delta) = {2\over 3} L^3\,,
\end{equation}
and
\begin{equation}
\sigma^{{\rm{\scriptscriptstyle LLA}}}_{\rm{\scriptscriptstyle PM}} = {Z^4\,\alpha^7
\over 3\,m_\mu^2} \, L^3 \,.
\end{equation}
The above result is in good agreement with the old result of \cite{Meledin} 
(see also Eq. (2.4) in review \cite{BGMS}). However, for the energies 
discussed in this paper the LLA does not provide sufficient precision. The 
ratio $G^{\rm{\scriptscriptstyle LLA}}/G$ is 1.5 for Pb at LHC and 2 for Au at RHIC.  
Hence, the LLA gives only a crude estimate for the energies discussed.

\section{ORTHODIMUONIUM PRODUCTION}
\label{secortho}

\begin{figure}[htb]
\centerline{\mbox{\epsfysize=4.0cm\epsffile{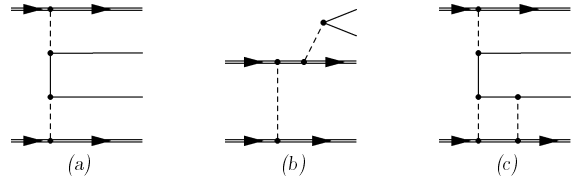}}}
\caption{\label{prod} Diagrams for two and three photon
production mechanisms of fermion pairs in relativistic heavy ion collisions.
In the case of dimuonium, the fermion pair is produced in a bound state.}
\end{figure}

\begin{figure}[htb]
\centerline{\mbox{\epsfysize=9.5cm\epsffile{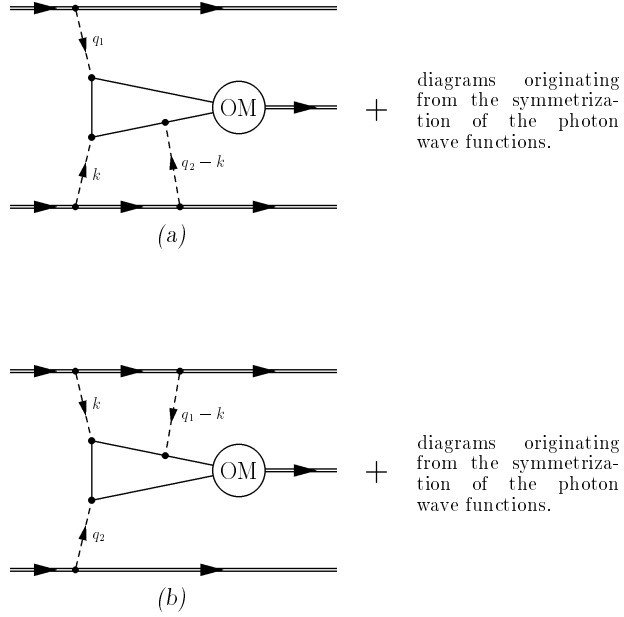}}}
\caption{\label{threephoton} Orthodimuonium production by a
three photon fusion process.}
\end{figure}

\begin{figure}[htb]
\centerline{\mbox{\epsfysize=9.5cm\epsffile{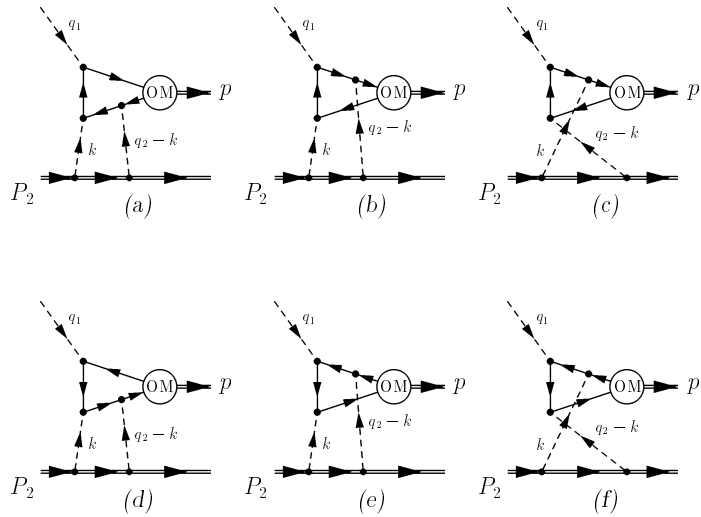}}}
\caption{\label{symm} Orthodimuonium production by a
three photon fusion process. $P_2$ denotes the nuclear momentum,
$p$ is the momentum of the dimuonium system.}
\end{figure}

Orthodimuonium can be produced by bremsstrahlung (the relevant diagram
is depicted in Fig.~\ref{prod}{\sl b}) and by three photon fusion
(see Fig.~\ref{prod}{\sl c}). For production processes induced by relatively 
light particles (like $e^+ e^-$ or $pp$ collisions) the three photon cross
section $\sigma_{3\gamma}$ corresponding to Fig.~\ref{prod}{\sl c} is 
suppressed by a factor $\alpha^2$ compared to the cross section for 
bremsstrahlung production, $\sigma_{\rm br}$. By contrast, for heavy ion 
collisions another parameter enters the calculation: the large nuclear mass 
$M$. Bremsstrahlung of heavy particles is suppressed by a factor $1/M^2$,
so we obtain 
\[
\sigma_{\rm br} \propto Z^6 \alpha^7/ M^2\,,
\]
whereas for three photon production there is no such suppression,
\begin{equation}
\label{order:3gamma}
\sigma_{3\gamma} \propto Z^6\alpha^9 / m_\mu^2\,.
\end{equation}
The ratio
\[
{\sigma_{\rm br} \over \sigma_{3\gamma}} 
\stackrel{<}{\sim} 
\frac{1}{\alpha^2}\,\left(\frac{m_\mu}{M}\right)^2 = 
\left\{\begin{array}{c l}
1/150 & \mbox{for RHIC, Au mode}\,, \\
1/190 & \mbox{for LHC, Pb mode}\,, \\
1/7   & \mbox{for LHC, Ca mode}
\end{array}\right.
\]
is small. Moreover, a more accurate estimate for CaCa
collisions at LHC decreases this ratio at least by a factor of three.
Therefore, the 3 photon production dominates in relativistic
heavy ion collisions.
In Fig.~\ref{prod}{\sl c} only one representative diagram for
three photon fusion is depicted. For a complete analysis, we need to take into
account two classes of diagrams, in which the single photon is emitted by 
either one of the nuclei (see Fig.~\ref{threephoton}{\sl a} and
Fig.~\ref{threephoton}{\sl b}). The corresponding cross section, which is 
proportional to the square of the amplitude for the particular 
processes, is given by
\begin{equation}
\label{dsigmaOM}
d\sigma_{\rm {\scriptscriptstyle OM}} = d\sigma_a + d\sigma_b + d\sigma_{\rm interf} = 
2 \, d\sigma_a
\end{equation}
because the interference term  $d\sigma_{\rm interf}$ disappears
after azimuthal averaging.
Thus we may restrict ourselves to an analysis of the cross section for the 
process in Fig.~\ref{threephoton}{\sl a}, denoted as $d\sigma_a$. 
To a very good approximation, this cross 
section can be expressed by the number of equivalent photons $dn_1$ emitted 
from one of the nuclei, given by Eq. (\ref{spectrum}), and the cross section 
for the process $\gamma A \to {\rm OM} + A$, denoted as $\sigma_{\gamma A}$,
\begin{equation}
\label{oneequiv}
d\sigma_a = dn_1 \, \sigma_{\gamma A}\,.
\end{equation}
We thereby assume that the incident photon in the process
$\gamma A \to {\rm OM} + A$ is a virtual photon in the framework of the 
equivalent photon approximation and thus exhibits a small virtuality
$Q^2 \ll \Lambda^2 < 4\,m_\mu^2$. 
Therefore, we neglect the virtuality of this incident photon in the
cross section $\sigma_{\gamma A}$. The 
subprocess $\gamma A \to {\rm OM} +A$ is described by the set of diagrams of 
Fig.~\ref{symm}. We calculate its cross section $\sigma_{\gamma A}$ in the 
region of large energies and relatively small transverse momenta 
$|{\bf p}_\perp|$ of the produced {\rm OM}. We have the kinematical conditions
(the subscript 'th' denotes the threshold value)
\[
s_\gamma = 2\,q_1 \cdot P_2 \gg s_{\rm th}=4\,m_\mu\,M \quad
\mbox{and} \quad |{\bf p}_\perp| \stackrel{<}{\sim} m_\mu\,.
\]
We note, that a loop integral has to be evaluated for this subprocess. Its 
contribution is rather different from what would be obtained within the 
standard equivalent photon distribution for the two remaining photons.

It is convenient to perform the calculations involved using the impact 
representation, which has been employed in QED and QCD for a number of 
processes with two photon or two gluon exchange (in the $t$-channel). 
More details on this approach are described in 
Refs.~\cite{LF,ChW,ginzburg1,ginzburg2}.
In this representation the amplitude $M_{\gamma A}$ which corresponds to
the whole set of diagrams of Fig.~\ref{symm} is written with an
accuracy $\sim m_\mu^2/s_\gamma$ in the form of a
two-dimensional integral over the transverse components of the
momentum of the virtual photon,
\begin{equation}
\label{mgamma}
M_{\gamma A} = i \int {d^2 {\bf k}_\perp \over (2 \pi)^2} \, 
{J_\gamma J_A \over
{\bf k}_\perp^2 ({\bf p}_\perp - {\bf k}_\perp)^2}\,.
\end{equation}
The impact factors $J_\gamma$ and $J_A$ correspond to the upper and lower 
block of the diagrams in Fig.~\ref{symm}. The diagrams in Fig.~\ref{symm} 
are regarded as being cut by the photon lines of the lower block, dividing 
the process into two partial virtual processes, $\gamma + \gamma \gamma
\to {\rm OM}$ (upper block) and $A \to \gamma\gamma +A$ (lower block). The 
impact factor $J_\gamma$ then corresponds to the virtual transition 
$\gamma + \gamma \gamma \to {\rm OM}$, and $J_A$ corresponds to the virtual 
transition in the lower block ($A \to \gamma\gamma +A $).
The impact factor $J_A$ for a charged point-like particle was found in 
\cite{LF,ChW} as $J_A= 4\pi \alpha Z^2$.  In our case we should take into 
account the shape of the nucleus and modify this impact factor according to 
\begin{equation}
J_A = 4\pi \alpha Z^2 \, F({\bf k}_\perp^2) \,
F(({\bf p}_\perp - {\bf k}_\perp)^2)\,.
\end{equation}
The impact factor $J_\gamma$ for the virtual transition 
$\gamma + \gamma\gamma \to {\rm OM}$ is similar to the impact factor for the 
virtual transition $\gamma + gg \to \Psi$ which was
introduced for the description of the hard diffractive process
$\gamma q\to \Psi q$ \cite{ginzburg2}. 
Adjusting for the different couplings and masses,
we immediately obtain
\begin{equation}
\label{jgamma}
J_\gamma = 4\pi \, \alpha^{9/2}\left[
{m_\mu^2 \over m_\mu^2 + {\bf p}_{1\perp}^2}-
{m_\mu^2 \over  m_\mu^2 + ({\bf p}_{1\perp} - {\bf k}_\perp)^2}
\right] \,
{\bf e }_{\gamma}{\bf e}_{\rm OM}^{*}\,.
\end{equation}
Here ${\bf p}_{1\perp} = {\bf p}_{2\perp} = 1/2\,{\bf p}_{\perp}$, and
${\bf e }_{\gamma}$ and $ {\bf e}_{\rm OM}$ are the polarization vectors for 
the initial photon and the final state OM.  
>From Eq.~(\ref{jgamma}) it follows that the helicity is conserved in the 
$\gamma \to {\rm OM}$ transition. Therefore, the OM is 
transversely polarized and is produced in two polarization states only
(not three states).

We finally obtain the cross section as 
\begin{equation}
d\sigma_{\gamma A} = 
Z^4 \,\alpha^8 \,
\left|  \Phi ({\bf p}^2_{\perp}) {\bf e}_\gamma
{\bf e}_{\rm OM}^* \right|^2 \, {d^2{\bf p}_{ \perp} \over m_\mu^4}\,,
\end{equation}
where $\Phi ({\bf p}^2_{\perp})$ is determined by an integral related to the 
amplitude $M_{\gamma A}$ given in Eq.~(\ref{mgamma}). 
$\Phi ({\bf p}^2_{\perp})$ can be written as
\begin{equation}
\label{defPhi}
\Phi({\bf p}^2_{\perp})= {1\over \pi} \int 
F\left(\frac{({\bf r} + {\bf n})^2 {\bf p}^2_{\perp}}{4} \right) \,
F\left(\frac{({\bf r} - {\bf n})^2 {\bf p}^2_{\perp}}{4} \right) \,
\frac{{\bf r}^2 -1}
{({\bf r} - {\bf n} )^2 ({\bf r}+ {\bf n})^2} \, {d^2 r \over
(1+\tau)(1+\tau \, {\bf r}^2)}
\label{3.13}
\end{equation}
where ${\bf r}$ is a two-dimensional vector with no physical dimension, over 
which the integration has to be performed. ${\bf n}$ is a unit vector 
defined by
\[
{\bf n}= {{\bf p}_{ \perp}\over |{\bf p}_{ \perp}|}\,,
\]
and $\tau$ is given by 
\[
\tau={{\bf p}_{\perp}^2\over 4\,m_\mu^2}\,.
\]
After integrating over the azimuthal angle of OM, summing over the
polarizations of the final state (OM spin states) and averaging over the
polarizations of the initial state (photon polarizations), we obtain
\begin{equation}
\label{sigmagammaA}
\sigma_{\gamma A} = B\, {\pi \, Z^4 \, \alpha^8 \over
m_\mu^2} \, {\Lambda^2 \over m_\mu^2}
\end{equation}
where the dimensionless constant $B$ follows from
\begin{equation}
\label{defB}
B = \int\limits_0^\infty ( \Phi ({\bf p}^2_{\perp}))^2 \,
\frac{d p^2_{\perp}}{\Lambda^2} =
\int\limits_0^\infty ( \Phi (\Lambda^2 u))^2 \, du
\end{equation}
and $\Lambda$ is the form factor scale 
defined in Eq.~(\ref{formapprox}). 

The value of the constant $B$ depends stronger on the shape of the form factor
than the corresponding quantity for paradimuonium. We used a
realistic form factor of the nuclear charge density 
$\rho({\bf r})$ for which we employed the model \cite{bohrmottelson}
\[
F({\bf k}^2) = \frac{1}{Z\,e} \,
\int d^3r\,e^{i {\bf k} \cdot {\bf r}} \, \rho({\bf r})\,,
\]
with
\begin{equation}
\rho({\bf r}) = {Z\,e\over N}\,{1\over 1 +
\exp\left((r-R)/a\right)}\,.
\label{defrho}
\end{equation}
The parameters are
\[
R = 1.18\,A^{1/3}\;{\rm fm} \,,\quad a = 0.53 \; {\rm fm}\,.
\]
$N$ is the normalization factor chosen such that
$\int d^3 r\,\rho({\bf r})=1$.

The evaluation of the constant $B$ is performed numerically on
{\sc IBM RISC/6000} workstations. Because the form factor
Eq.~(\ref{defrho}) is a function of ${\bf k}^2\,R^2$ with $R
\propto \Lambda^{-1}$, the constant $B$ has
the same value for all nuclei considered in this paper.  We
obtain
\begin{equation}
\label{resB}
B = 0.85\,.
\end{equation}

It is useful to consider the sensitivity of the result on the choice
of the form factor. With the approximate form
factor given in Eq.~(\ref{formapprox}) the function $\Phi$ is calculated in
the Appendix \ref{app:b} analytically (with an additional
approximation $\tau=0$ in the denominators of
Eq.~(\ref{defPhi})). Further evaluation results in $B=0.93$,
this value is in fair agreement with the exact value from
Eq.~(\ref{resB}).

Because the cross section of the subprocess
$\gamma A \to {\rm OM} + A$ (cf.~Eq.~(\ref{sigmagammaA})) 
is energy independent in the discussed
limit, the remaining integration of Eq.~(\ref{oneequiv}) is in
fact an integration over the equivalent photon spectrum
only. Let $\omega_2$ be the total energy of
both exchanged photons in Fig.~\ref{threephoton}. Then we have
as in the previous section $4\,\omega_1\,\omega_2 = 4\,m_\mu^2$
(due to four-momentum conservation and the kinematics of the
process) and $\omega_2\stackrel{<}{\sim}\Lambda \gamma $ (due to
the nuclear form factor). Thus a lower limit for $\omega_1$ is
\begin{equation}
\omega_1 >  {m_\mu^2 \over \omega_{2{\rm max}}} \approx {m_\mu^2 \over
\Lambda \gamma}\,.
\label{3.5}
\end{equation}
The upper bound in this integration can be set  to $\infty$
due to the fast decrease of the equivalent photon
spectrum at large energy.  We obtain
(using the notation $\delta = m_\mu/(\Lambda \gamma)$
introduced previously)
\begin{equation}
\label{sigmaA}
\sigma_a = {Z^2 \alpha \over \pi} \, H(\delta) \,\sigma_{\gamma A}\,,
\end{equation}
where
\begin{equation}
H(\delta)= \int\limits_{m_\mu^2 / (\Lambda \gamma)}^\infty
{d\omega_1\over \omega_1} \,
f\left({\omega_1 \over \Lambda \gamma }\right) =
\int_{\delta^2}^\infty {dx\over x}\, f(x)=\left\{
\begin{array}{c l}
57 & \mbox{for RHIC, Au mode},\\
202 & \mbox{for LHC, Pb mode}, \\
247 & \mbox{for LHC, Ca mode}.
\end{array}
\right.
\end{equation}

Finally, the cross section for the OM production is equal to 
(cf.~Eqs.~(\ref{order:3gamma},\ref{dsigmaOM},\ref{sigmagammaA},\ref{sigmaA})),
\begin{equation}
\label{sigmaOM1}
\sigma_{\rm {\scriptscriptstyle OM}} = 2\,{Z^6\,\alpha^9 \over
m_\mu^2} \, {\Lambda^2 \over m_\mu^2} \,B\,H(\delta )\,.
\end{equation}
The numerical values are
\begin{equation}
\sigma_{\rm {\scriptscriptstyle OM}} = 10^{-30}\, {\rm cm}^2 \times
\left\{\begin{array}{c l}
0.021     & \mbox{ for RHIC, Au mode}\,,\\
0.089     & \mbox{ for LHC, Pb mode}\,,\\
0.000069  & \mbox{ for LHC, Ca mode}\,.
\end{array}\right.
\label{sigmaOM2}
\end{equation}
The ratio for the production cross section for the ortho and para states 
is given by
\begin{equation}
{\sigma_{\rm {\scriptscriptstyle OM}}\over 
\sigma_{\rm {\scriptscriptstyle PM}} } =
4\, \left({Z\alpha\Lambda \over m_\mu}
\right)^2\, B\, {H(\delta) \over G(\delta)}\,
= \left\{\begin{array}{c l}
0.144 & \mbox{for RHIC, Au mode}\,, \\
0.066 & \mbox{for LHC, Pb mode}\,, \\
0.010 & \mbox{for LHC, Ca mode}\,.
\end{array}\right.
\end{equation}
Hence, we expect predominantly a production of para states in relativistic
heavy ion collisions.

\section{ESTIMATE OF THE BACKGROUND}
\label{secback}

The dimuonic atoms are neutral
systems produced by a number of photons which are approximately on-shell
and collinear to the colliding ions. So the angular spread of the dimuonia with 
respect to the beam-axis is of the order of ${\cal O}(\gamma^{-1})$.
The rapidity of the DM particles will be correspondingly 
high. Therefore the DM systems will not be observed directly by any detector with a 
low rapidity coverage. 

The dominant decay channels of DM are $\gamma\gamma$ (for 
PM) and $e^+e^-$ (for OM, cf.~Eq.~(\ref{decay})). 
Dimuonium could be observed via detection 
of these decay products. We will investigate in this Section the influence of 
three sources of background on the prospective measurements,
\begin{itemize}
\item photon and electron-positron background originating
from hadronic processes in the interaction region,
\item free electron-positron pair production 
shadowing the signal from the decaying orthodimuonium and
\item photon pair production by the two colliding nuclei
shadowing the signal from the decaying paradimuonium.
\end{itemize}
The above sources of background are expected to affect mainly
the signal of those dimuonium atoms which decay in or near the interaction 
region of the heavy ion collision.

First we consider photon and electron-positron background originating from 
inelastic hadronic processes in the interaction region. This background will 
affect the signal from both OM and PM atoms. In these inelastic 
processes, one or both of the nuclei dissolve to some extent. One can roughly 
divide these processes into two classes.  The first are mainly hadronic 
processes, where the two nuclei collide and the strong interaction takes 
effect. The cross section for this class can be estimated as 
$\sigma_{AA} \approx 4\,A^{2/3}\sigma_{pp}$, which for the nuclei under 
investigation is in the range of five to seven barn. The second type is a 
photodissociation process caused by an energetic photon emitted from a larger 
distance by one of the nuclei. It induces nuclear reactions in the other 
nucleus on impact. The cross section of this photodissociation process
depends crucially on the type of the colliding ions. One finds cross sections 
of roughly 85 barn for RHIC in the Au mode, 200 barn for LHC in the Pb 
mode and 3 barn for LHC in the Ca mode \cite{KGS}. In the following we list 
the approximate luminosities per bunch crossing ${\cal L}_b$ and the 
corresponding probability of hadronic events per bunch crossing
${\cal P}_h = \sigma_h \, {\cal L}_b$, where the hadronic cross section 
$\sigma_h$ is a combination of the purely hadronic and the photo--dissociation
cross section. We obtain
$$
{\cal L}_b= \left\{\begin{array}{c } 2.2
\times 10^{19}\,{\rm cm}^{-2} \,,\\
3.75 \times 10^{20}\,{\rm cm}^{-2} \,,\\
10^{23}\,{\rm cm}^{-2} \,,
\end{array}\right.\quad
{\cal P}_h\approx
 \left\{\begin{array}{c l}
0.004 & \mbox{for RHIC, Au mode}\,, \\
0.075 & \mbox{for LHC, Pb mode}\,, \\
1 & \mbox{for LHC, Ca mode}\,.
\end{array}\right.
$$
By virtue of this figures we may conclude, that for Au and Pb the hadronic
event rates are small enough to see DM-production in anticoincidences with 
the production of additional hadrons. In contrast this seems to be quite an 
impossible task for the Ca mode at LHC. 

A second, significant source of background for the decay of the OM 
is caused by the production of free $e^+e^-$ pairs by the two nuclei.
In order to estimate this effect, we consider the $e^+e^-$ pair production via
the standard two-photon mechanism. The cross section of this process, 
$\sigma_e$, is estimated with the well-known Racah-formula (see \cite{BGMS} 
for details). We obtain $\sigma_e\approx 35\,000\,{\rm barn}$ for RHIC and 
$\sigma_e\approx 225\,000\,{\rm barn}$ for LHC. This is orders of magnitudes larger
than the production cross section for the DM. A remedy for this problem
might be a precise determination of the invariant mass of the
electron-positron pair. The production cross section for an $e^+e^-$ pair 
having an invariant mass near $2\,m_{\mu}$ with mass spread $\Delta m$ is
calculated using Eq.~(\ref{dsigmaPM}) with the replacement of 
$\sigma_{\gamma\gamma}$ by the cross section for the process
$\gamma\gamma\to e^+e^-$:
\begin{equation}
\Delta\sigma(A_1\,A_2 \to A_1\,A_2\,e^+\,e^-)=
{(Z\alpha)^4  \over \pi \, m_{\mu}^2}\,G(\delta )\,
\left( \ln{4m_\mu^2\over m_e^2}-1\right){\Delta m\over m_\mu}\,
\approx 1.8\times 10^7{\Delta m\over m_\mu}\,
\sigma_{\rm {\scriptscriptstyle PM}}.
\label{eebckgr}
\end{equation}
Because the orthodimuonium production rate is comparably low, even a realistic
mass resolution of $1\,{\rm MeV}$ would not fit our goal to distinguish 
the background from the signal of the 
orthodimuonia. Hence, it seems to be a very difficult task
to observe the orthostate of dimuonium in heavy ion collisions.

This situation is different for the parastate.
The main non--hadronic background to the the deacy of the PM atoms is
the two photon production process $A_1\,A_2 \to A_1\,A_2 + \gamma\,\gamma$. 
It can be 
described as the radiation of two (virtual) photons and subsequent light by 
light scattering (via an electronic loop). The cross section for this process 
is given by Eq.~(\ref{dsigmaPM}) with the replacement of 
$\sigma_{\gamma\gamma}$ by the cross section of light by light scattering. 
It is five orders less than $\sigma_e$ and for invariant masses $>200\,{\rm MeV}$
of the photon pair it drops by five more orders. Nevertheless, the
total free photon pair production cross section for all energies
greater than  $200\,{\rm MeV}$ is still
larger than the production cross section for PM by a
factor of $\approx\alpha^{-1}$. To improve the situation one might again
try to fix the invariant mass $m$ as precisely as possible.
In general, a relative precision $\Delta m/m$ at $m^2 \approx 4\,m_\mu^2$  
will lead to a signal-to-noise ratio of the order of 
$\alpha^{-1}\,\Delta m/m$. 
\begin{equation}
\Delta\sigma(A_1\,A_2 \to A_1\,A_2\,\gamma\,\gamma)= 0.95\,
{Z^4\,\alpha^6 \over 2 \, m_{\mu}^2}\,G(\delta )\,
{\Delta m\over m_\mu}
\approx {1\over \alpha}{\Delta m\over m_\mu}\,
\sigma_{\rm {\scriptscriptstyle PM}}.
\label{ggbckgr}
\end{equation}
So the signal to background ratio becomes about 0.75 for mass resolution 
$\Delta m =1$ MeV. This corresponds to a determination 
of the invariant mass of the decay 
products of PM with a precision of roughly $5\times10^{-3}$. By contrast, 
for OM, a determination of the invariant mass of the
electron-positron pair to an accuracy of the 
order of $10^{-7}$ would be necessary in order to reach a comparable 
signal-to-noise ratio.

Another possibility to further improve the situation is to take into account
only highly relativistic dimuonium systems, which decay outside
the nuclear collision region.
The electron-positron or photon pair (for OM and PM, respectively)
is produced outside the interaction region. 
DM atoms decay after travelling a typical decay length is $\ell
\approx\varepsilon/(2\,m_{\mu})c \tau$ (see Table~\ref{restable}). This
distance is increased for the excited dimuonia. The
opportunity to observe excited dimuonium states in this approach
will be better than that for the ground state despite the smaller production
rate.  The task left in this picture is to reconstruct the vertices of the
decays, which necessitates a vertex detector.

\section{RESULTS AND DISCUSSION}
\label{secres}
 
We have analyzed the production of bound states consisting of a muon
and an antimuon in relativistic heavy ion collisions. The analysis 
of the parastate production was performed for the 
dominating two photon process. The two photon
approximation describes the production of the parastate with
an accuracy of $(1\div 2)\%$. A novel three photon mechanism
for the production of the orthostates was developed. The
accuracy of this approximation is $(6\div 12)\%$. 
The theoretical uncertainty of our results is primarily
due to multiphoton processes. Other sources of uncertainty 
like the dependence  on the nuclear form factor 
or corrections to the equivalent photon spectrum, have been analyzed in 
detail. They are on the level of one precent. 
Because multiphoton processes enhance the production rate,
our results should be regarded as a lower bound on the total production.

We obtained numerical results for the 
dimuonium production at the new heavy ion
colliders RHIC and LHC. The results for all colliders and for a set of
atomic states are presented in Table~\ref{restable}. 
In Table~\ref{restable} we also consider the properties
of the atomic 2P dimuonic state which is produced in atomic transitions from 3S
and 4S\footnote{The
results are obtained according to the treatment of recoil
effects in Ref.~\cite{Fried}. The atomic transitions in heavy
fermionium has been discussed in Ref.~\cite{Moffat}. The DM
spectrum is considered in detail in Ref.~\cite{nlo}.}.

The dimuonic atoms travel, after production, with small angular spread 
along the beam axis. Therefore they are detectable by their decay products only.
In general the extraction of a signal from the experiment will be easier 
for paradimuonium than for orthodimuonium. The reasons are twofold,
\begin{itemize}
\item the total production cross section is much larger and
\item the background is significantly reduced.
\end{itemize}
As it has been shown in Sec.~\ref{secback}, the 
photon pair background shadowing the 
paradimuonium signal is roughly five orders of magnitude smaller than
the free electron-positron pair background shadowing the signal from  
orthodimuonium. Additionally, the total production cross sections for para 
states are larger than those for ortho states by a factor of 10 to 100, 
depending on the collider and the nucleus used (see Table \ref{restable}).
We expect a favourable signal-to-noise ratio 
for the parastate if the energy of the photon 
pair can be determined with a precision of roughly $1\,{\rm MeV}$.

\section*{ACKNOWLEDGEMENTS}

I.~F.~G., S.~K. and V.~G.~S. are grateful to the Max-Planck-Institut
f\"ur Physik komplexer Systeme for warm hospitality. U.~D.~J. wishes
to thank DAAD for continued support. This work was supported
in part by the grants INTAS--93--1180ext and the grant RFBR
96-02-19079 from the Russian Foundation for Basic Research. The work of 
U.~D.~J., F.~K. and G.~S. was additionally supported by GSI (Darmstadt), BMBF 
and DFG. D.~Ivanov, V.~Ivanov, K.~Melnikov and 
G.~Baur are gratefully acknowledged for stimulating discussions.

\appendix
\section{PARADIMUONIUM LIFETIME}
\label{app:a}

\begin{figure}[htb]
\centerline{\mbox{\epsfysize=5.5cm\epsffile{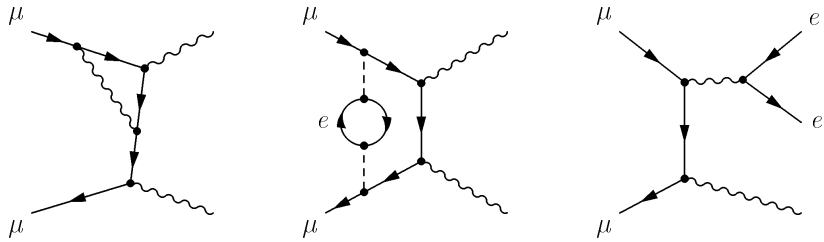}}}
\caption{\label{pmnlo} Typical NLO corrections to the
PM decay rate.}
\end{figure}

\begin{figure}[htb]
\centerline{\mbox{\epsfysize=5.5cm\epsffile{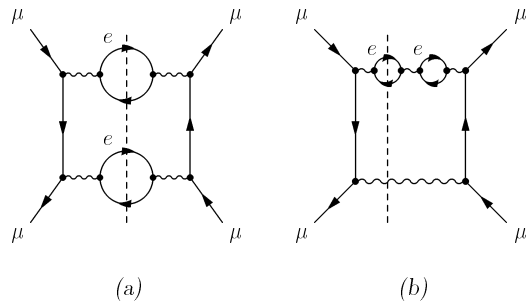}}}
\caption{\label{pmnnlo} 
Feynman diagrams for the double logarithmic 
NNLO corrections to the PM decay rate.}
\end{figure}

We consider briefly the lifetime of the parastates 
of dimuonium. Because of the higher production rate of parastates, 
this is of interest in the context of possible experiments.
The leading term is caused by the 
$\gamma\gamma$ decay,
\begin{equation}
\Gamma^{(0)}(n{^1}S_0) = \frac{\alpha^5 \, m_\mu}{2\,n^3}\,.
\end{equation}
Typical contributions to the
next-to-leading order (NLO) corrections are depicted diagrammatically 
in Fig.~\ref{pmnlo}. The corrections are evaluated in \cite{nlo} as
\begin{equation}
\Delta \Gamma^{{\rm {\scriptscriptstyle NLO}}}(1 {^1}S_0) = 
4.79\,\frac{\alpha}{\pi}\,  \Gamma^{(0)}(1 {^1}S_0)
\end{equation}
and
\begin{equation}
\Delta \Gamma^{{\rm {\scriptscriptstyle NLO}}}(2 {^1}S_0) =
4.65\,\frac{\alpha}{\pi}\,  \Gamma^{(0)}(2 {^1}S_0)\,.
\end{equation}
The next-to-next-to-leading order (NNLO) corrections include the
large logarithmic factors $\ln(1/\alpha)$ and $\ln^2(m_\mu/m_e)$. We consider
here these logarithmic terms. The $\ln(1/\alpha)$ term is of the same form
as for parapositronioum \cite{khriplo},
\begin{equation}
\Delta \Gamma^{{\rm {\scriptscriptstyle NNLO}}}_1(n {^1}S_0) = 
2\, \alpha^2\, \ln\left( \frac{1}{\alpha}\right)\,\Gamma^{(0)}(n {^1}S_0) \,.
\end{equation}
The double mass ratio logarithm does not have an analogy in parapositronium.
The relevant Feynman diagrams are depicted in Fig.~\ref{pmnnlo}. We
obtain the result,
\begin{equation}
\Delta \Gamma^{{\rm {\scriptscriptstyle NNLO}}}_2(n {^1}S_0) = 
(1 + 2)\, \frac{4\,\alpha^2}{9 \, \pi^2} \, 
\ln^2\left( \frac{m_\mu}{m_e}\right) \, \Gamma^{(0)}(n {^1}S_0)\,.
\end{equation}
The terms $(1 + 2)$ originate from the diagrams in Fig.~\ref{pmnnlo}{\sl a}
and Fig.~\ref{pmnnlo}{\sl b}, respectively.
The calculation of the double logarithmic corrections is done here
by the evaluation of the imaginary part of the diagrams in Fig.~\ref{pmnnlo}.
The real part of these diagrams
contributes to the hyperfine structure in higher order. 
The final results for the lifetimes are
$\tau(1{^1}S_0) = 0.59504(22) \times 10^{-12} \, {\rm s}$ and
$\tau(2{^1}S_0) = 4.7619(17) \times 10^{-12} \, {\rm s}$.

\section{DEPENDENCE OF THE ORTHOSTATE PRODUCTION ON THE FORM FACTOR}
\label{app:b}

Employing the approximate form factor given in Eq.~(\ref{formapprox}),
we evaluate the integral Eq.~(\ref{defB}) analytically.
First we observe that the predominant contribution to $B$ is caused by the
region where $p^2_{\perp}/ \Lambda^2 < 1$.
In this region $\tau = p^2_{\perp}/( 4 \,m_\mu^2) < 0.1$. Therefore we 
may put $\tau =0$ in Eq.~(\ref{defPhi}). We integrate
\begin{equation}
\label{phianalytic}
\Phi(\Lambda^2 \, u)= {1\over \pi} \int {({\bf r}^2 -1) \over
({\bf r}- {\bf n} )^2 ({\bf r}+ {\bf n})^2} \, {d^2 r \over
[1+ (u/4) ({\bf r}- {\bf n})^2]  [1+ (u/4) ({\bf r}+ {\bf n})^2]}\,.
\end{equation}
The relation
\[
{\bf r}^2 -1 = {1\over 2} \,
\left[({\bf r}+ {\bf n})^2 + ({\bf r}- {\bf n})^2 - 4\right]
\]
proves to be useful for a simplification of the integrand. 
We can present the integral in Eq.~(\ref{phianalytic}) 
in the form
\begin{equation}
\label{phiasI}
\Phi (\Lambda^2 \, u) = - (2+a_1) \,I_{11} - 2 \,I_{22} + (4+a_1) \,I_{12} \,.
\end{equation}
The integrals $I_{ij}$ are defined as
\[
I_{ij} = {1\over \pi} \int { d^2 r \over
[({\bf r}+ {\bf n} )^2 +a_i]\,[({\bf r}- {\bf n})^2 +a_j]}\, ,
\]
where $a_1 = 4/u$. The auxiliary parameter
$a_2 = \epsilon \to 0$ is introduced in order to
regularize divergences in intermediate calculations.
For the integrals $I_{ij}$ we obtain after elementary integration
\[
I_{11} = {1\over \sqrt{1+a_1}} \, 
\ln{{1+\sqrt{1+a_1}\over\sqrt{a_1} }}\,, \;\;\;
I_{22} = {1\over 2} \ln{{4\over \epsilon}}\,, \;\;\;
I_{12} = {1\over 4+a_1}\, \ln{{(4+a_1)^2 \over a_1 \epsilon}} \,.
\]
Substituting these expressions into Eq.~(\ref{phiasI})
we obtain the result 
\begin{equation}
\Phi (\Lambda^2 \, u) = \ln{{(1+u)^2 \over u}} - 
{4+2u \over \sqrt{4u+u^2}}\,
\ln{{\sqrt{4+u}+\sqrt{u}}\over 2}\,, 
\end{equation}
in which the dependence on the auxiliary parameter $\epsilon$ vanishes.
Using this method we obtain a result of $B = 0.93$ in contrast to
$B=0.85$ with the exact form factor (cf.~Eq.~(\ref{resB})).

\begin{table}[htb]
\begin{center}
\begin{tabular}{c|c|c|c|c|c}
&&&&& \\[-1ex]
Collider & Nucleus & Nuclear & Luminosity  & Lorentz & Bunch\\[1ex]
& & charge & ${\cal L}$  & factor  & length  \\[1ex]
& &  ($Z$) &  [cm$^{-2}$s$^{-1}$] & (${\gamma}$) &  [cm] \\[1ex]
\hline
&&&&&\\[-1ex]
RHIC & AuAu & 79 & $2 \times 10^{26}$ & 108  & $ 12$   \\[1ex]
LHC  & PbPb & 82 & $3 \times 10^{27}$ & 2980 & 7.5     \\[1ex]
LHC  & CaCa & 20 & $4 \times 10^{30}$ & 3750 & $ 7.5$  \\[1ex]
\end{tabular}
\end{center}
\caption{\label{colli} Experimental parameters of RHIC and LHC which
must be taken into account for the production of dimuonium.  The
bunch length of $7.5\,{\rm cm}$ for the CaCa-channel at LHC is
an estimate.}
\end{table}

\begin{table}[htb]
\begin{center}
\begin{tabular}{ccc|cc|c|c|c}
\multicolumn{5}{c|}{}
&\multicolumn{2}{c}{}\\[-1ex]
\multicolumn{5}{c|}{ATOMIC STATE PROPERTIES}&
\multicolumn{2}{c}{ESTIMATED PRODUCTION PER YEAR}\\[1ex]
\hline
&&&&&&&\\[-1ex]
Atom & State & $J^{PC}$ & $c\tau$ & Decay &  RHIC & LHC & LHC
\\[1ex]
 &  &  & [cm] & mode &Au--Au&Pb--Pb&Ca--Ca \\[1ex]
\hline
&&&&&&&\\[-1ex]
PM & $1{^1}S_0$ & $0^{-+}$ & 0.0178
               & $\gamma\gamma$       &   310 & 40000 & 260000 \\[1ex]
PM & $2{^1}S_0$ & $0^{-+}$ & 0.143
              & $\gamma\gamma$         &   40  & 5000 & 33000\\[1ex]
PM & $3{^1}S_0$ & $0^{-+}$ & 0.483
              & $\gamma\gamma$  &   12  & 1500 &
9800 \\[1ex]
PM & $4{^1}S_0$ & $0^{-+}$ & 1.14
              & $\gamma\gamma$&   5  & 630 & 4100
\\[1ex]
PM & $8{^1}S_0$ & $0^{-+}$ & 9
              & $\gamma\gamma$&  --  &  79 & 520
\\[1ex]
PM & $10{^1}S_0$ & $0^{-+}$ & 18
              & $\gamma\gamma$&  --  & 40  & 260
\\[1ex]
\hline
&&&&&&&\\[-1ex]
OM & $1{^3}S_1$ & $1^{-{-}}$ & 0.0538
              & $e^+e^-$   &   43  & 2700 & 2800 \\[1ex]
OM & $2{^3}S_1$ & $1^{-{-}}$ & 0.430
              & $e^+e^-$         &  5 & 330 & 340 \\[1ex]
OM & $3{^3}S_1$ & $1^{-{-}}$ & 1.45
              & $e^+e^-$       & 2   & 100 & 100 \\[1ex]
\hline
&&&&&&&\\[-1ex]
PM & $2{^1}P_1$ & $1^{+{-}}$ & 0.462
              & $1{^1}S_0\;  \gamma$ &   --    & 60 & 400 \\[1ex]
\end{tabular}
\end{center}
\caption{\label{restable} Main properties of atomic states
of dimuonium and their estimated production at LHC and RHIC per
year (running time per year in our calculation is $10^7$ s). The {\em decay
mode} given here is the dominant mode which is most important for 
the detection.}
\end{table}

\end{document}